# Fabrication of the preferentially orientated $LiNbO_3$ nanoparticles and their ferroelectric behaviors

Cheng Li and Shining Zhu[a]

*Department of Physics, National Laboratory of Solid State Microstructures,*

*Nanjing University, Nanjing 210093, People's Republic of China*

Long Ba[a]

*State Key Laboratory of Bioelectronics, School of Biology Science and Medical Engineering,*

*Southeast University, Nanjing 210096, People's Republic of China*

Zhenlin Luo

*National Synchrotron Radiation Laboratory, University of Science and Technology of China, Hefei*

*230029, People's Republic of China*

(00l)-Oriented $LiNbO_3$ ultra-thin film was fabricated on Pt/Ti/$SiO_2$/Si substrate using pulsed laser deposition. The film could shrink into crystalline state nanoparticles that still kept (00l) preferential orientation as being annealed in the oxygen atmosphere at 500 °C for 2 hours. These $LiNbO_3$ nanoparticles exhibited the ferroelectricity, showing asymmetrical hysteresis loop which might originate from the large internal field at the interface. These nanoparticles could be used for further studying on the size effects and the ferroelectricity of nanocrystal.

a) Electronic mail: zhusn@nju.edu.cn and balong@seu.edu.cn

Domain engineering in bulk Lithium Niobate (LiNbO$_3$) single crystal, of increasing importance for the nonlinear optics [1] and electro-optic Bragg gratings [2], has been intensively investigated during the last decades [3,4]. With respect to downscaling potential for integrated optics, many studies especially based on the technique of Piezoresponse Force Microscopy (PFM), have been accumulated on the switching behavior of nanosize ferroelectric domain in LiNbO$_3$, such as the nanodomain in single crystal [5,6] and polycrystalline thin film [7]. In these situations mentioned above, however, the different electrical and mechanical boundaries as well as the random grain crystallization orientation complicate the analysis of the ferroelectric behavior. Hence it is important to elucidate the mechanism of the polarization dynamics in isolated ferroelectric nanoparticles with known orientation. In this aspect, LiNbO$_3$, as one kind of *uniaxial* ferroelectric crystal [8], has advantage over other ferroelectrics, such as BaTiO$_3$ and PbTiO$_3$, because only 180º domain exists in it. This enables LiNbO$_3$ nanocrystal as a unique sample to study the size effect and other size related properties in ferroelectric material.

It should be noted that, to realize the epitaxial growth, niobium doped SrTiO$_3$ (Nb: STO) [9] or La$_{0.5}$Sr$_{0.5}$CoO$_3$ (LSCO) [10] are widely used as the bottom electrode, while platinum coated tip is usually utilized as the movable top electrode. However, this configuration with intrinsic asymmetry complicates the interpretation of the ferroelectric behavior of nanocrystal. Hence, in this letter, we fabricated the preferential (001) orientation LiNbO$_3$ nanoparticles on the platinized substrate. We utilized PFM to study the surface topography and ferroelectricity. These nanoparticles exhibited the asymmetrical hysteresis loop attributed to the existence of a large internal field. The internal field might mainly originate from a interface layer, referred to as a nonswitching layer in the letter, between the electrode and nanoparticles.

In order to fabricate nanoparticles, the ultra-thin $LiNbO_3$ film about 20 nm thickness on the commercial $Pt/Ti/SiO_2/Si$ substrate was firstly deposited by pulsed laser deposition (PLD) with a KrF excimer laser (248 nm, ~3-4 $J/cm^2$, 3 Hz). The deposition was carried out in the oxygen atmosphere of 30 Pa for 3 min, while the deposition temperature was fixed at 600 $^oC$. Details of the similar preparation process have been reported in other literature [11]. After deposition, the whole film was annealed in the oxygen atmosphere at 500 $^oC$ for 2 hours. During the process, the discrete $LiNbO_3$ nanoparticles were formed with size ranged from 30 nm to 200 nm.

The domains in these nanoparticles were visualized by a PFM system (vecco Multimode V, Digital Instruments) with internal lock-in amplifier, and the local switching behaviors were detected by configuring a function generator (Agilent 33120A) and a DSP lock-in amplifier (SR830 Stanford Research Instrument). In order to optimize signal to noise ratio, the frequencies were set at 10.8 kHz and 12.6 kHz, for vertical and lateral PFM, respectively. Pt coated silicon cantilevers (Ultrasharp NSG11, NT-MDT) was used, with spring constant 5 N/m, and curvature radius of tip about 10 nm.

To acquire the local polarization vector component, PFM was operated in contact and lateral force mode with internal lock-in amplifier, referred to as vertical PFM (VPFM) and lateral PFM (LPFM). Briefly, ac voltage was applied between the probing tip and the grounded nanoparticle, which deformed due to the converse piezoelectric effect. The vertical and lateral piezoresponse, sensed by the four quadrant photodiode, spitted into amplitude and phase signal by the lock-in amplifier, could be used to present the local polarization direction. This method has been described as Vector PFM [12].

Eliseev *et al* have proved the relationship between the Vector PFM signal and the crystallographic orientation [13]. Taking into account the LiTaO$_3$ piezoelectric strain tensor as well as the general transformation from the laboratory coordinate system to the crystal coordinate system, it can be obtained that the out-of-plane piezoresponse is maximum, and both of the orthogonal in-plane piezoresponse are zero when the polar axis, with the same direction of C-axis (i.e. (00l) orientation), is parallel to the surface normal. Owing to the similar trigonal structure in LiNbO$_3$, this approach could also be used to identify the (00l) orientation of LiNbO$_3$ nanoparticle.

As a reference experiment, we first measured a sample with densely particles distribution. Figure.2 (a)-(c) exhibit the sample's surface topography, VPFM phase and amplitude images. The dark and bright areas in VPFM phase image represent negative and positive domain (i.e. the component of polarization vector orients upward and downward), respectively. While, Figure 2 (d)-(f) are the surface topography, LPFM phase and amplitude images of the same area, however, most part does not present obvious contrast. This is consistent with the measurement of X-ray diffraction (XRD) (see Fig.1), indicating these nanoparticles are preferentially (00l) orientated. However, some nanoparticles (solid circle) still have in-plane orientation component, showing the contrast in both VPFM and LPFM phase image. So instead of the average orientation information, Vector PFM could reveal local crystallographic orientation.

The topographic image of well-separated nanoparticles is shown in Fig.3 (a). Note that during the annealing process, the interdiffusion, defect accumulation and the aggregation may lead to the decay of original preferential orientation [7]. However, most particles still maintain the film's initial (00l) orientation. In this situation, PFM is required to specify the orientation of individual nanoparticle. As shown in Fig.3 (f) and (e), however, apart from the obvious VPFM phase contrast, there are LPFM responses detected at the edge of the three particles without significant signal in the centre. Similar phenomena were observed by Peter *et al.* who concluded that in-plane piezoresponse at the perimeter is due to the topography of the nanoparticle [14]. Hence these nanoparticles still keep the (00l) orientation.

In general, nanoparticle should exhibits many different phase contrasts, bright and dark, which depend on the orientation of $180^{\circ}$ multidomain. In Fig.3 (b), however, the phase image of particle A shows uniform contrast, whereas that of particle B shows two distinct areas. This indicates particle A is a single domain nanocrystal and particle B contains two different ferroelectric domains, one is positive and the other is negative. It is proposed that due to the surface effect [15] or the depolarization field effect [16], there is a critical size for ferroelectric nanocrystal. Two or more domains can co-exist in the same particle only when its size is larger than the critical size [17]. If further decreasing the size, the ferroelectricity would decay. In Fig.3 (b), the lateral size of particle C, which is smaller than the above ones, is about 100 nm. Its image contrast is not so clear as that of A and B nanoparticle, which implies the decrease of piezoresponse as the size scaling down. Similar phenomena were also observed during the transition from ferroelectric phase to superparaelectric phase in other kinds of ferroelectrics [17, 18].

To quantitatively analyze the ferroelectricity of individual LiNbO$_3$ nanoparticle, a typical isolated nanoparticle with (00l) orientation, lateral size of ~70 nm and the height of ~40 nm was selected for the local hystersis loop measurement. Accounting for the high convolution by parasitic effect in such small size [19], the piezoresponse hysteresis loop and corresponding phase loop are both exhibited in Fig.4, indicating the polarization direction could still be switched by external electric field down to size of 70 nm. The maximum piezoresponse value ~2 pm/V exhibits decrease compared with ideal bulk value ~6 pm/V [8], revealing size effect on these ferroelectric nanoparticles. However, the exact critical size for the transition to superparaelectric phase requires further investigation based on statistical analysis. We also observed large asymmetry in the loop, i.e. a significant horizontal shift of the hysteresis loop along the electric field axis. The asymmetry suggests that there exists a large internal field in LiNbO$_3$ nanoparticle along (00l) orientation. From Fig.4, we can attain the magnitude of the coercive field to be about 60 kV/mm and the internal field $E_{int}$ is about 40 kV/mm, according to the formula $E_{int}=(E_f+E_r)/2$, where $E_f$ is the switching field for forward poling, and $E_r$ for reverse poling. We are aware of that the shift of loop could be generated by different kinds of bottom and top electrodes. It is ascribed to the difference in work function steps at two interfaces of particle-electrode [20]. In this work, however, both of the top and bottom electrodes are platinum, so the effect should vanish due to the symmetrical electrode configuration.

There are two possible explanations for the origin of internal field. The first one is due to domain pinning induced by point defects [4]. These defects originate from the composition deviation from the stoichiometry, including the oxygen vacancies and lithium vacancies, which are ascribed to the relative low oxygen atmosphere during the film fabrication and anneal processes [7]. However, the internal field due to nonstoichiometric point defects in previous works [21], is only 2~3 kV/mm internal field with 21 kV/mm coercive field, the ratio is much smaller than the one in nanoparticle. Hence nonstoichiometric point defects could not be the main reason, though it plays an important role in the completed picture. The second possible reason involves built-in electric field at the interface between Pt bottom electrode and $LiNbO_3$ nanoparticles. The field, attributed to interdiffusion process during the annealing and the defect dipole accumulation, is high enough to polarize a thin layer in the vicinity of the electrode. It is the self polarized layer that leads to asymmetrical hysteresis loop. Such large internal field was also detected by W. Wu *et al* [22]

The microscopic origin of interface layer that is not switchable under the external field needs the further investigation. Here we can utilize the incomplete 180° switching model [23] to calculate its thickness, which is not switchable under the external field. The (00l) orientation and the *uniaxial* poling property of these nanoparticles guarantee the validity of this simple model. Relative hysteresis loop

offset Ω can be calculated:

$$\Omega = (d^+ - d^-)/(d^+ + d^-), \qquad (1)$$

where $d^+$ and $d^-$ are the remanent piezoelectric coefficients obtained from the measured piezorespose hysteresis loop. Assuming the piezoresponse signal d is proportional to the normal component polarization [24], $d^+$ and $d^-$ would be proportional to the volume of positive domain and negative domain, respectively. According to the model, the thickness of the nonswitching layer is calculated about 10 nm, which is consistent with the estimated values in the range from 10-15 nm based on other models [25,26].

In conclusion, the preferentially (00l) oriented LiNbO$_3$ nanoparticles were fabricated by PLD and a post-anneal process, the crystallographic orientation was confirmed by the XRD and the Vector PFM. It is proposed that the interface layer between Pt electrode and LiNbO$_3$ nanoparticle plays important role in forming a high internal field, which leads to asymmetrical piezoresponse hysteresis loop. These LiNbO$_3$ nanoparticles could be considered as a suitable sample to study the size effect of domain and other defect related physical properties of ferroelectric nanocrystal.

Acknowledgements

L. Ba grateful acknowledges the Natural Science Foundation of China for partial support of this work through grants 10576008, 10774022. Authors thank Prof. Shoujun Xiao of School of Chemistry and Chemical Engineering, Nanjing University for AFM assistance.

Figure captions

FIG. 1. XRD pattern of LiNbO$_3$ particles prepared on Pt/Ti/SiO$_2$/Si substrate. The peaks labeled with circles indicate the (00l) reflection of LiNbO$_3$. Note that, to improve the signal/noise ratio, the crystallographic observation by XRD was carried out on a sample with denser LiNbO$_3$ particles, which are deposited for 10 min.

FIG. 2. Surface topography (a,d), vertical piezoresponse phase and amplitude (b,c), and lateral piezoresponse phase and amplitude (e,f) for the LiNbO$_3$ sample. The phase images provide the polarization orientation information, and the amplitude images illustrate the region with the strong (white) and weak (black) piezoelectric response.

FIG. 3. (a,d) Surface topography, (b,c) vertical phase and amplitude and (e,f) lateral phase and amplitude for well seperated nanoparticle. Note that the out of plane response is relatively uniform, whereas the signal of in-plane is small in the center and high at the edge of the particle.

FIG. 4. Both the piezoelectric response hysteresis and phase loop measured on an individual LiNbO$_3$ nanoparticle are shown. For all figures, (■) d$_{33}$ for the piezoresponse hysteresis loop, i.e. Acosθ signal, where A is piezoresponse amplitude and θ is phase, whereas (●) d$_{33}$ for the phase loop

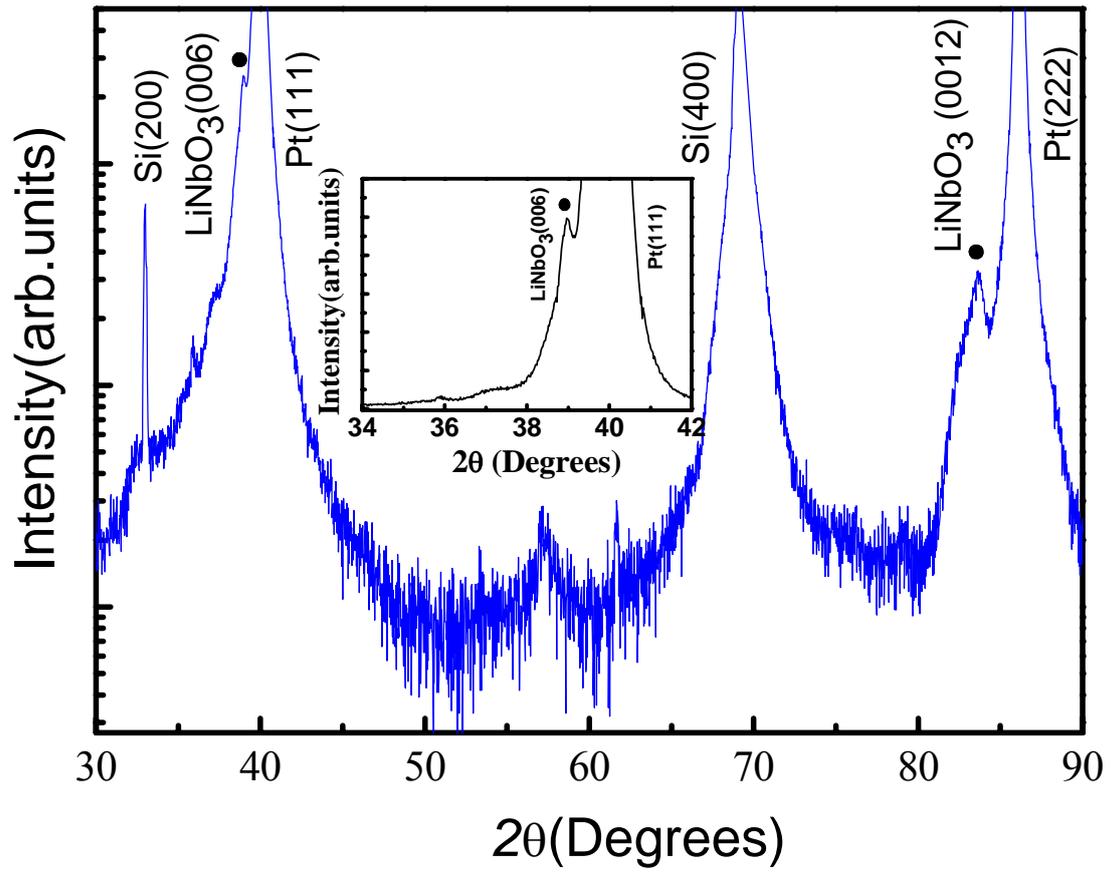

FIG. 1.

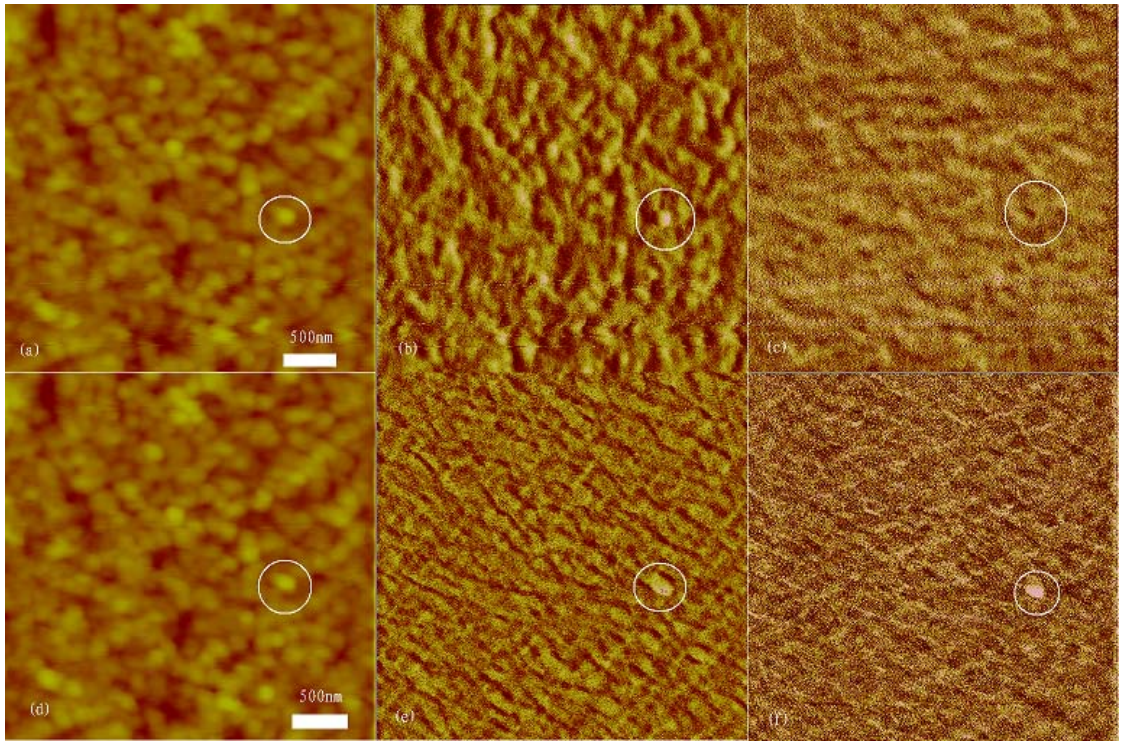

FIG.2.

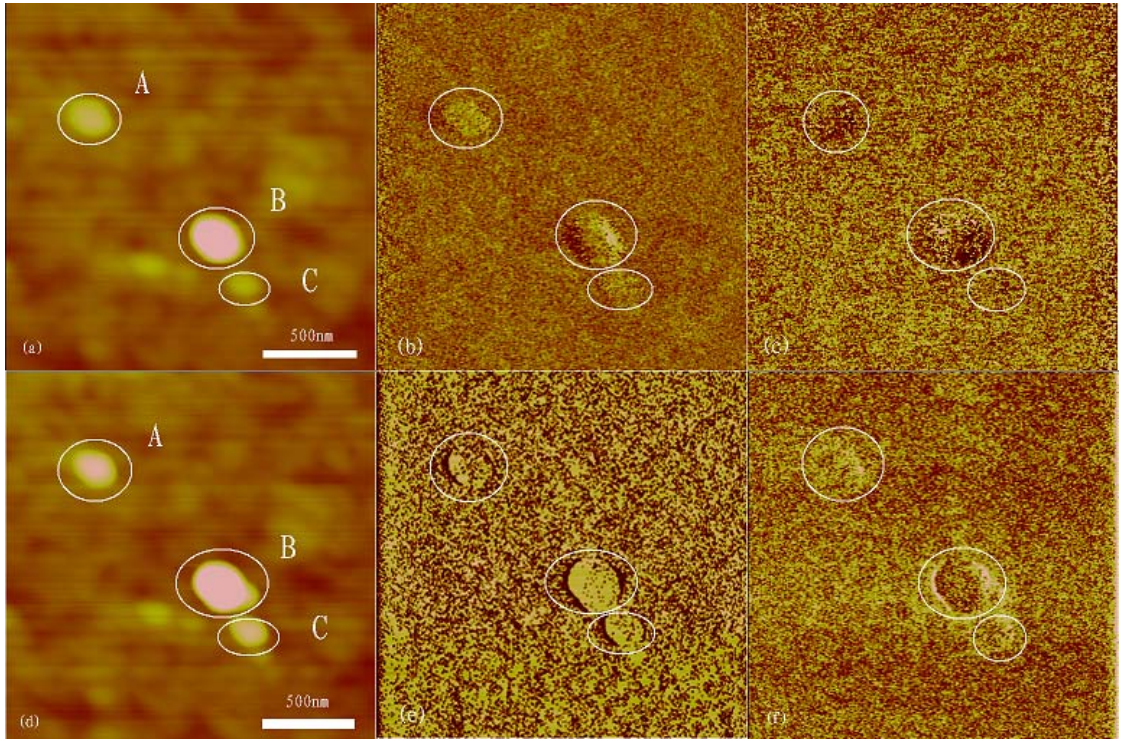

FIG. 3.

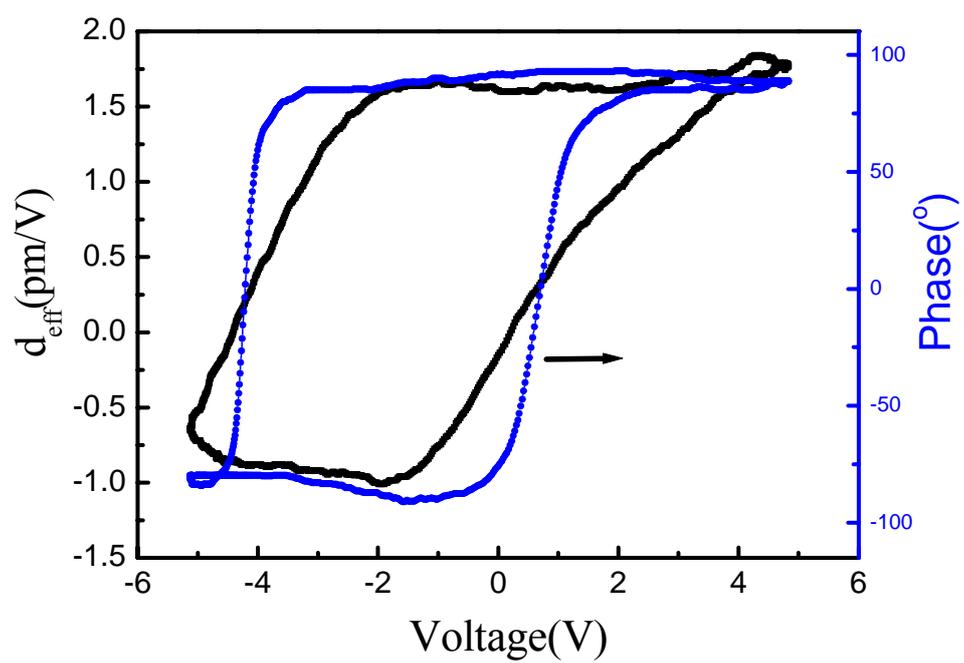

FIG.. 4.